\documentclass[twocolumn,journal]{IEEEtran}
\usepackage{amsmath}
\usepackage{graphicx}
\usepackage{cite}
\usepackage{epstopdf}
\usepackage{balance}
\usepackage{gensymb}
\usepackage{color}
\usepackage{lipsum}
\usepackage{float}
\usepackage{epstopdf}
\usepackage[mathcal]{eucal}
\usepackage{subfiles}
\usepackage[T1]{fontenc}
\usepackage[unicode=true,
 bookmarks=true,bookmarksnumbered=true,bookmarksopen=true,bookmarksopenlevel=1,
 breaklinks=false,pdfborder={0 0 0},backref=false,colorlinks=false]
 {hyperref}
\hypersetup{pdftitle={Your Title},
 pdfauthor={Your Name},
 pdfpagelayout=OneColumn, pdfnewwindow=true, pdfstartview=XYZ, plainpages=false}

\newcommand{\ignore}[1]{}

\makeatletter

 \let\oldforeign@language\foreign@language
 \DeclareRobustCommand{\foreign@language}[1]{%
   \lowercase{\oldforeign@language{#1}}}

\usepackage[caption=false,font=footnotesize]{subfig}

\makeatother

\begin{document}

\bstctlcite{IEEEexample:BSTcontrol}

\title{Empirical Effects of Dynamic Human-Body Blockage in 60 GHz Communications}

\author{Christopher Slezak, Vasilii Semkin\textsuperscript{*}, Sergey Andreev, Yevgeni Koucheryavy, and Sundeep Rangan
\thanks{C.~Slezak and S.~Rangan are with NYU WIRELESS, NYU Tandon School of Engineering, Brookyln, NY 11201, US.}
\thanks{V.~Semkin, S.~Andreev, and Y.~Koucheryavy are with Tampere University of Technology, Tampere 33101, Finland.}
\thanks{V.~Semkin, is also with Peoples' Friendship University of Russia (RUDN University), 6 Miklukho-Maklaya St, Moscow, 117198, Russian Federation.}
\thanks{\textsuperscript{*}A part of this work has been completed during the research visit of Dr. Sc. (Tech.) Vasilii Semkin to NYU WIRELESS, US, supported by Jorma Ollila grant.}
\thanks{This work was supported in part by NSF grants 1302336, 1564142, and 1547332, NSA, NIST, SRC, Sony, and the affiliate members of NYU WIRELESS. The work of V. Semkin was supported in part by the Finnish Cultural Foundation (Suomen Kulttuurirahasto) and in part by ”RUDN University Program 5-100.” This work was also supported in part by the Academy of Finland (project PRISMA).}
}

\maketitle

\begin{abstract}
The millimeter wave (mmWave) bands and other high frequencies above 6~GHz have emerged as a central
component of Fifth-Generation (5G) cellular standards to deliver high data rates and ultra-low latency.
A key challenge in these bands is blockage from obstacles, including the human body.
In addition to the reduced coverage, blockage can result in highly intermittent links where the
signal quality varies significantly 
with motion of obstacles in the environment.  The blockages have widespread consequences throughout 
the protocol stack including beam tracking, link adaptation, cell selection, handover and congestion
control.  Accurately modeling
these blockage dynamics is therefore critical for the development and
evaluation of potential mmWave systems.  In this work, we present a novel spatial dynamic channel sounding system
based on phased array transmitters and receivers operating at 60 GHz.  Importantly, the sounder can measure multiple directions
rapidly at high speed to provide detailed spatial dynamic measurements of complex scenarios.
The system is demonstrated in an indoor home-entertainment type setting with multiple moving blockers.
Preliminary results are presented on analyzing this data with a discussion of the open issues towards developing
statistical dynamic models.

\end{abstract}


\IEEEpeerreviewmaketitle{}


\section{Introduction}

\normalsize

Since the inception of cellular communications systems, providers have faced an ever-increasing demand for data.
In recent years,  the impending capacity crunch has
forced the many stakeholders 
to explore higher frequency bands where more spectrum is available and harness them for mobile communication~\cite{Shafi2017}. These extremely high frequencies include the millimeter wave bands, which formally comprise the range of 30 to 300~GHz.  These bands,
along with other frequencies above 6~GHz, are now considered the primary vehicle for enabling the extremely high throughputs of bandwidth-hungry applications and services in fifth-generation (5G) mobile networks.


However, the design and deployment of mmWave communication systems
present several unique challenges arising from
the relatively unfavorable radio wave propagation at these frequencies. 
Most significantly, mmWave signals are highly susceptible to \emph{blockage}.
Compared to signals transmitted at lower (e.g., microwave) frequencies, mmWave transmissions simply cannot 
penetrate many common objects and building materials.
Similarly, the human body, the subject of this work, can typically result in 20~dB of attenuation as demonstrated later in this article --
much greater than at lower frequencies.

\ignore{This vulnerability to blockage is perhaps the most serious concern
for the deployment and feasibility of future mmWave systems.  
Most obviously, blockage significantly reduces the range
and coverage of cells and access points.  In outdoor urban micro-cellular deployments, for example, 
measurement studies have predicted that
cell sizes will likely be limited to 200~m or less, with significantly higher
levels of outage than conventional micro-cellular deployments today~\cite{RanRapE:14}. }

In addition to loss in coverage, blockage results in another equally challenging problem:
\emph{channel dynamics}.  Due to blockage, mmWave signal strength can vary rapidly with any motion relative to
obstacles in the environment.  For example, moving behind buildings, changes in the orientation of
a handset or the location of a body or hand blocker may all cause fast and significant large-scale fading. 
These dynamics have a wide-ranging impact throughout the protocol stack including
beam tracking, link adaptation, channel feedback at the physical layer, as well as Transmission Control Protocol (TCP) and end-to-end
application performance. While there have been extensive studies of blockage in the mmWave bands, many of these works have
been limited to static scenarios.  
For example, there are numerous
studies for penetration loss measurements of various materials, as well as
measurement campaigns in outdoor settings for deriving statistical path loss and coverage models
with fixed receiver locations. 

Understanding channel dynamics in realistic time-varying blockage scenarios is more complicated.  
Most importantly, mmWave transmissions are inherently directional.  To overcome the high isotropic path loss,
mmWave signals will be transmitted and received in narrow beams formed by electrically steerable 
antenna arrays.  Due to scattering and reflections, 
there are often multiple viable communication paths in addition to the line-of-sight (LOS) links.
Hence, if a current mmWave path is suddenly blocked, ongoing communication can be steered
to an alternate path from the same serving cell or an alternate cell to maintain the Quality of Service/Experience
~\cite{fengLinkBlockage}.
MmWave blockage models must therefore describe the joint dynamics of these paths to properly assess 
such procedures including
beam tracking, multi-connectivity, and handover.  

\ignore{In this article, we begin
with a review of blockage modeling techniques and measurement studies that have been developed as part of the WiFi (802.11)
and cellular (3GPP) standardization '
efforts~\cite{gustafson2011HumanShad60G}.  The models can be derived analytically
(via diffraction theory)~\cite{gapeyenko17MobBlockers} and have been experimentally verified. 
However, we argue that the models may be difficult to apply to complex scenarios with large numbers of moving obstacles.}

In this article, we begin
with an overview of blockage modeling techniques which view blockage from a Radio Frequency (RF) theory perspective \cite{gustafson2011HumanShad60G}, and those which seek to develop statistical models and methodologies for simulations involving mmWave blockage \cite{jacob11ad60GModel}. These statistical models often rely on ray-tracing based studies instead of measured data due to the difficulty of performing these measurements.

To ensure that there is no gap between these blockage models and the true statistical nature of human blockage, NYU WIRELESS has developed a 60 GHz measurement system using two phased arrays. Importantly, the 
system can measure the channel characteristics in multiple directions almost simultaneously. \ignore{enabling
a complete spatial time-varying profile of the channel} We report measurements in an indoor, home entertainment-type setting with multiple 
moving blockers and discuss some preliminary efforts aimed at deriving statistical characterizations of 
blockage dynamics from these measurements.

\ignore{The system can be applied to complex blockage scenarios with motion of the transmitter, receiver and obstacles.  
.  We also discuss , including a low-rank tensor decomposition method for 
identifying channel components.}

\ignore{These limitations motivate our main contribution, which is the development of a 
novel spatial dynamic channel measurement system based on phased arrays. }


\section{Overview of Blockage Models}
\label{sec:overview}

\normalsize 

\begin{figure}[!ht]
	\centering{}
	\includegraphics[scale=0.62]{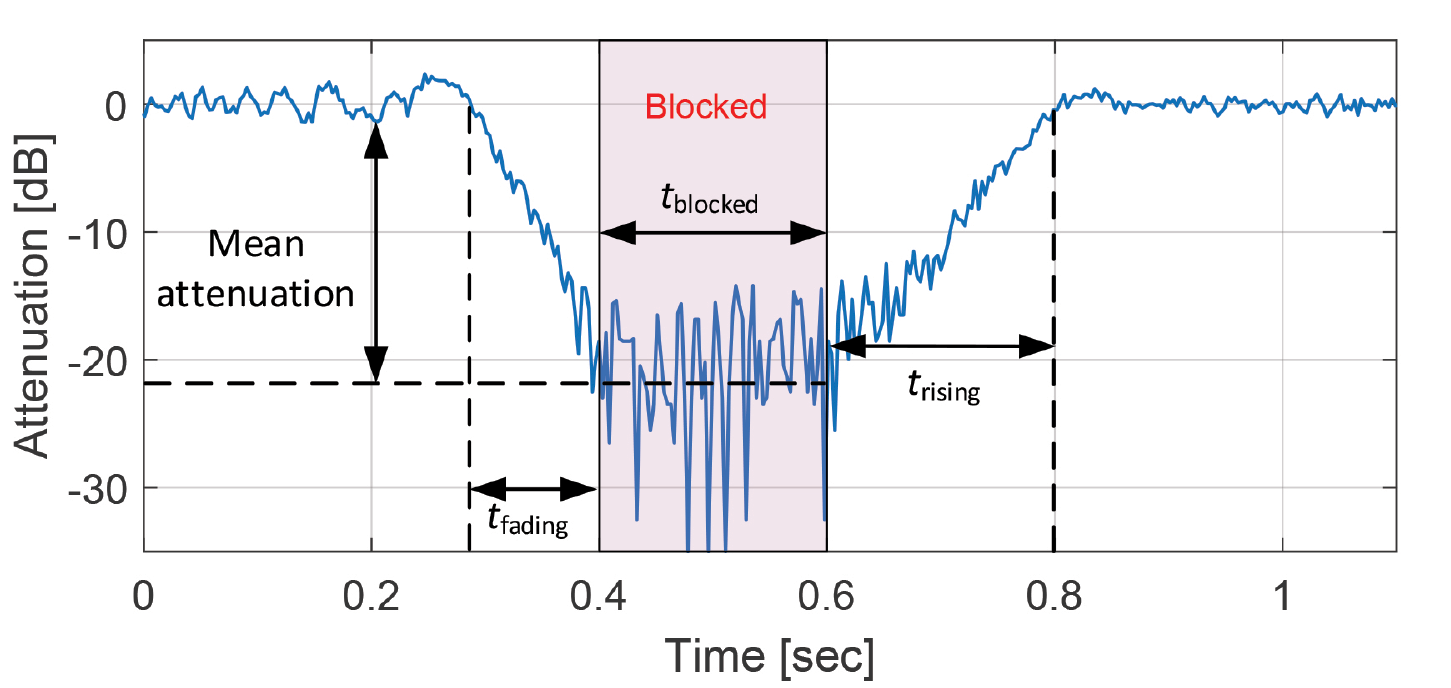}
	\caption{Example blockage event obtained from our measurements.}
	\label{fig:blockage}
\end{figure}

Before describing blockage models, it is useful to consider a typical human-body blockage event.  Fig.~\ref{fig:blockage} presents a trace of such an event
from our measurements discussed in detail below. The figure shows the attenuation versus time for the LOS component.  Blockage models seek to characterize the statistics
of various key parameters such as the attenuation level, the duration of the blockage, as well as the fading and rise time.

We can see that the impact of human-body blockage on the radio link can be severe.  In the trace of Fig.~\ref{fig:blockage}, there is a more than 20~dB
attenuation for a duration of $t_{\text{blocked}} = 0.2$ seconds.
Furthermore, before and after the link is blocked, there are regions $t_{\text{fading}}$ and $t_{\text{rising}}$, where the signal strength first decreases as the blocker begins to obstruct the LOS path and then rises as the blocker moves away. These effects alone can influence the quality of a mmWave connection. For 5G wireless systems requiring low latency, such blockage durations are critical and must be mitigated by beam steering antenna arrays. If the LOS signal is blocked, the link can be maintained by switching to an alternative path.




\subsection{Blockage Studies Extending RF Theory}

\ignore{Models for understanding mmWave blockage can be broadly classified into those addressing the statistics of blockage and others seeking extensions/confirmation of RF theory.}Many early papers on mmWave blockage focused on 60~GHz due to the standardization efforts behind IEEE 802.11ad and beyond \cite{gustafson2011HumanShad60G}, \cite{jacob11ad60GModel}. More recently, with the increasing interest in mmWave for 5G there have been a growing number of contributions that targeted other frequencies, such as 28 and 73~GHz. The authors of~\cite{rappaportSmallScalePropagation} reported a series of measurements performed with a single-input single-output (SISO) channel sounder operating at 73~GHz. Two horn antennas were pointed directly at each other to yield a channel that was dominated by a single LOS component, where a human blocker walked through the link at varying distances between the transmitter (TX) and the receiver (RX). The resulting data was utilized to verify and extend the well-known double knife edge diffraction (KED) model to allow its use with directional antennas at mmWave frequencies. In this model, the attenuation due to a screen between the TX and the RX is computed analytically, requiring precise knowledge of the TX, RX, and screen locations. In addition to the double KED model, other models have received attention such as multiple KED and uniform theory of diffraction. As argued in~\cite{peter11blockmeas}, alternative blockage models may under- or overestimate attenuation; thus, measurements are needed in order to accurately characterize human-body blockage. \ignore{One common approach for human-body blockage modeling is through ray-tracing based simulations which are calibrated using relatively scarce measurement data. Then, this model derived from the simulations is introduced into the channel model.}

\ignore{However, any practical application of the aforementioned models requires knowledge of the geometry and exact motion of a blocker. Obtaining such descriptions constitutes a modeling effort on its own, particularly in complex scenarios with large numbers of blockers. In contrast, we seek to construct our human-body blockage model by relying on measurements that employ beam steering and therefore obtain full spatial information about the radio channel which can further be applied in arbitrary scenarios.}

\ignore{Based on this data and the results reported in the following sections, reliable channel models can be constructed and utilized in the corresponding standardization activities.}

\subsection{mmWave Blockage Models for Simulation Purposes}

Geometry based blockage models are a very common approach which allows the simulation to capture the effects of blockers moving throughout the environment. As noted earlier, the question is whether or not the locations and behavior of these blockers can be accounted for accurately. Despite these concerns, these types of models are widespread and can be very useful in the appropriate circumstances. 

The 3rd Generation Partnership Project (3GPP) channel model for frequencies from 0.5 to 100 GHz \cite{3GPP38901} includes blockage as an add-on feature. The blockers are distributed randomly within the environment in a way that is dependent upon the scenario under simulation. To evaluate attenuation due to blockage, there are two methods available. The first is a simplified model, which remains computationally efficient, while the second is a more intensive but realistic calculation that models each blocker as a screen and then follows the KED theory to calculate the attenuation due to blockage. 

The METIS initiative \cite{METISRevised} also utilizes geometry-based blockage modeling with screens similar to 3GPP. All the considered screens are vertical and oriented perpendicular to the line connecting the TX and the RX. For sparse distributions of blockers, METIS proposes to sum the losses due to each of the screens, whereas in denser distributions a more sophisticated technique based on the Walfisch-Bertoni model is employed to determine the effects of multiple blockers.

In \cite{wangStochasticGeometry}, the authors perform an extensive simulation of mmWave blockage in a vehicle-to-infrastructure (V2I) scenario. Vehicles are randomly dropped into an environment which has base stations located at street corners. The effects of large blocking vehicles such as buses and trucks are considered and important figures such as coverage probability are derived analytically and verified via numerical simulations. Using this framework, the authors were able to uncover some interesting insights, such as the fact that blockage can be considered insignificant for V2I scenarios with multiple-lanes as opposed to a single-lane roadway.

\ignore{In contrast to these geometry-based methods which require simulations to physically model the locations of blockers within the environment, other formulations can be developed which model the distribution of blockers in a more abstract fashion. In~\cite{maccartneyRapidFadingCrowds}, measurements were conducted for a 73~GHz mmWave point-to-point link with both TX and RX at ground-level on either side of a walking path. Human blockers traveled along the walking path to create blockage events on this link, and the collected data was used to create a Markov model that captured the probability that the link will transition between unblocked and blocked states. This type of modeling is attractive because it can be simulated with very little computation, while its simplicity makes it easy to compare results from different sets of measurements.}



\subsection{Common Measurement Methods}
\label{sec:measMethods}

One common method to study dynamic blockage is by employing a combination of physical measurements and ray-based simulations. The measurement systems typically fall into several categories: (i) fixed horn or patch antennas possibly with mechanical steering, (ii) quasi-omni antennas, and (iii) MIMO measurement systems with a limited number of antennas. Horn antennas provide excellent spatial filtering, which makes them well suited for many types of channel measurements. However, this spatial filtering is a double-edged sword when it comes to studying channel dynamics where one would like to simultaneously measure the channel across all multipath components that are available. Because human blockage is a process that unfolds over timescales on the order of milliseconds, it is impossible to rotate a horn antenna quickly enough to perform measurements over all of these paths. 

This problem is partially resolved by utilizing antennas with an omnidirectional pattern, where it is indeed possible to simultaneously measure blockage of different multipath components. However, this capability comes at the cost of spatial information. Multipath components are identified solely by the different times at which they arrive at the receiver. \ignore{With an accurate model of the environment, it is possible to work backwards and identify the likely paths by which these signals propagated, but this process is time consuming and difficult to scale.} As will be discussed later, phased array systems can bridge the gap between these different types of measurement systems.\ignore{and have been used to study blockage on a relatively coarse time scale \cite{qualcomm5GNRBlockage}.}

Another consideration for a blockage measurement campaign is whether or not the blockers should be aware of the experiment and follow predefined paths, or whether they should be people moving throughout an environment in a completely unbiased manner. For measurements seeking to verify RF theory it makes sense to have the blockers follow simple paths for repeatability \cite{rappaportSmallScalePropagation}, whereas for measurements that are being used to understand the statistics of blockage it is preferable to have the blockers be unaware of the experiment \cite{collogneHumanActivity}. A potential third option is to have blockers follow predefined paths that are designed to mimic the motion of ordinary people.

\section{Blockage Measurement Methodology}

\subsection{Experimental Setup}
\label{sec:systemDescription}

To address the limitations of common measurement techniques discussed earlier, NYU WIRELESS has constructed a setup that utilizes SiBeam 60~GHz phased antenna arrays at both the TX and RX~\cite{slezak60GHzBlockage}. 
Each transceiver array consists of 12 steerable TX elements and 12 steerable RX elements.  The array has a steerable range of approximately $\pm 45 \degree$ in the azimuth plane with a peak directional gain
of 23~dBi.  This system enables directional measurements to be performed similar to studies that relied on mechanically rotated horn antennas, but is additionally capable of switching from one angle of arrival/departure (AoA/AoD) to another with microsecond latency --
fast enough to keep up with human blockage events. The SiBeam phased array can electronically alter its pointing angle by applying a weight vector which defines the relative phases applied to the elements of the antenna array.

A crucial drawback of phased arrays is that the radiation patterns that can be achieved are often of much poorer quality than those of horn antennas. For example, in \cite{x60Paper} the authors present simulated antenna patterns for the same arrays that are used in this work. They note that the half-power beamwidths are typically around $30 \degree$ and in some cases the mainlobe and sidelobe are nearly at the same level, particularly as the mainlobe gets farther away from the boresight. Luckily, different multipath components can still be discriminated via the transmission of a wideband sequence which separates different arriving multipath components in the channel. This system can be thought of as a middle ground between horn antennas which provide excellent spatial resolution but can only measure a single multipath component, and omnidirectional measurements which provide no spatial information and therefore rely entirely on delay to distinguish different multipath components.

\begin{figure}[!ht]
	\centering
	\includegraphics[scale=1.02]{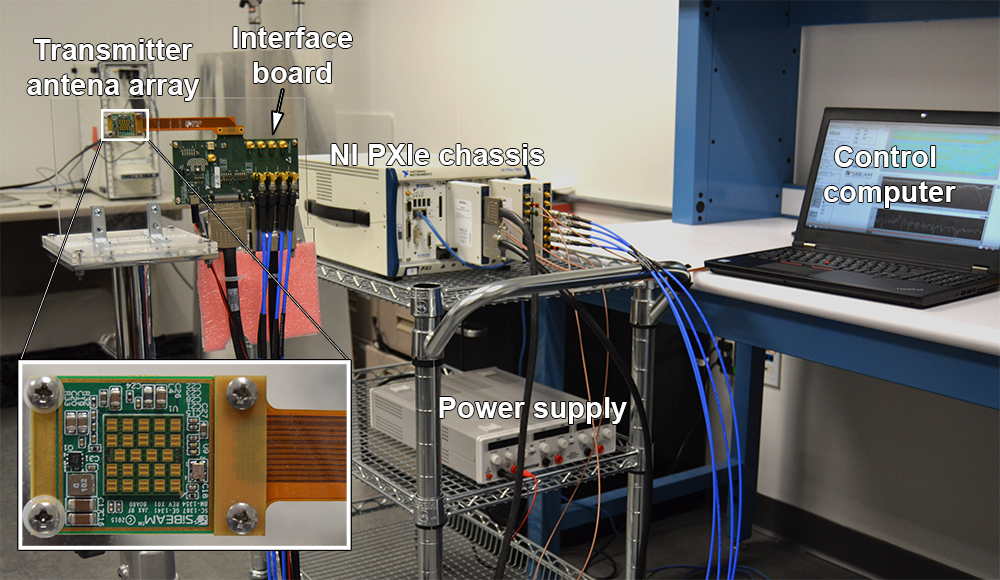}
	\caption{Photograph of the transmitter side of our measurement setup, including the SiBeam array which forms the RF portion and the PXIe chassis for baseband processing.}
	\label{fig:equip}
\end{figure}

Our system uses two codebooks of 12 predefined antenna weight vectors that allow for the TX and RX array to be scanned over a dozen of distinct angles of arrival/departure with an approximate range of coverage of $\pm45\degree$ in the azimuth plane. Each array is supported by a National Instruments PCI eXtenstions for Instrumentation (PXIe) chassis. Fig.~\ref{fig:equip} displays the TX component of the system. The chassis contains several Field Programmable Gate Arrays (FPGAs) as well as Input/Output (I/O) modules that allow them to interface with the array. These FPGAs implement all of the necessary signaling to control the arrays during a measurement as they rapidly scan over the azimuth plane. Cables connecting the TX and the RX chassis ensure that these remain synchronized throughout an experiment.

\subsection{Measurement Procedure}

In each measurement, we collect hundreds of thousands of complex-valued channel impulse response (CIR) samples over a period of about 5 seconds. To perform a scan over all of the possible combinations of AoA and AoD, the TX will first apply one weight vector to the array, and the RX will cycle through its 12 weight vectors and acquire a CIR for each. The TX will then advance to the next weight vector in its codebook, with the RX again cycling through its entire codebook. A single scan of all 144 AoA/AoD combinations takes less than one millisecond, and this scan is repeated approximately every 3 milliseconds. 

The duration of a single scan is short relative to the speed of a moving blocker; thus, we can assume that all CIRs captured during one scan experience the same blockage profile. In other words, one can conclude that these CIRs are captured `almost' simultaneously. \ignore{This rapid repetition of scans is also fast enough to ensure that the position of blockers traveling at typical walking speeds do not change significantly between scans.} This is a crucial feature of our system which separates it from other similar setups that can only examine blockage over a single path at a time.

\subsection{Indoor Blockage Measurements}
\label{sec:blockageMeasurements}
\begin{figure}[!ht]
	\centering
	\includegraphics[scale=0.7]{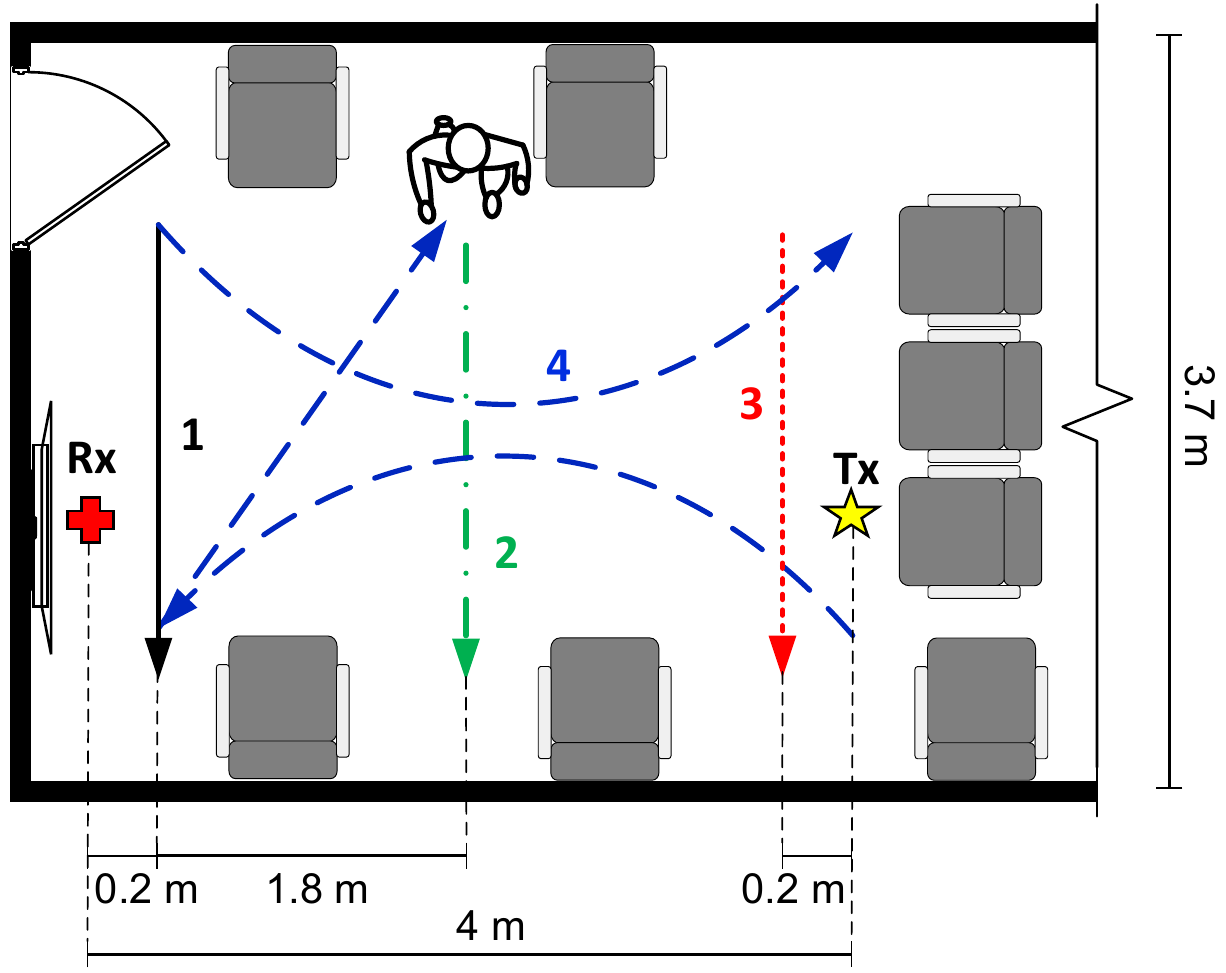}
	\caption{Layout of the measurement environment and trajectories of moving blockers.}
	\label{fig:plan}
\end{figure}

To evaluate the effects of human-body blockage in a characteristic in-home streaming/gaming scenario, a measurement campaign was conducted in a room on the NYU campus with furnishings that mimicked a typical living room. The TX was located at a height of 1~m beneath a wall-mounted TV to model a set-top box or similar device located as part of an entertainment center, while the RX was located 4~m away in front of a couch to represent a seated user. While the measurement system was continuously scanning, human blockers moved around the room at typical walking speeds following the paths indicated in Fig.~\ref{fig:plan}.

Trials 1, 2, and 3 featured a single human blocker walking perpendicular to the LOS path between the TX and the RX. This simple blocker motion serves as a useful baseline and provides a good opportunity to compare with other studies such as the one in \cite{maccartneyRapidFadingCrowds}. Trial 4 comprised multiple blockers attempting more `natural' paths as if they were entering/exiting the room or moving from one seat to another. This trial was repeated with the number of blockers varied from 1 to 3 in order to examine the effects associated with the density of blockers. \ignore{The paths were chosen to model one person exiting the room and two others moving to different locations while the user remains seated on the couch directly opposite the television.} Finally, all trials were performed with the arrays turned $\pm60\degree$ in the azimuth plane. This allowed the experiments to better capture the non-LOS (NLOS) component reflected off the walls of the room. 

This setup was chosen because it represents a scenario of significant interest (wireless virtual reality being one example of short-range, high throughput, and low latency communications) but it is certainly not the only situation where this measurement system can be useful to understand mmWave blockage. For example, additional measurement campaigns could focus on in-home streaming with relays located throughout the home, or in a crowded cafe with a very high density of blockers that are in almost constant motion.

\section{Empirical Results of Human Blockage}

\subsection{Interpreting Measurement Results}
\label{sec:dataTensor}

The results of our measurements take the form of a very large 4-way tensor, which essentially represents a 4-dimensional array. This tensor is indexed by the delay (measured in samples of the CIR), the RX antenna weight vector index, the TX antenna weight vector index, and the time at which the scan was acquired. It is possible to use this measurement data ``as-is'' by taking a measurement trace and using it to analyze the performance of a beamtracking algorithm or an entire mmWave communications system which employs antennas similar to those used in these measurements. As we will see however, developing statistical models from this measurement data can benefit from some additional processing of the measurement files.

The multidimensional nature of the measurement data, as well as its large volume, may make it cumbersome to analyze and interpret the results. A simpler approach is to reduce the order of the tensor by performing interleaving (commonly referred to as partial unfolding) of the RX antenna weight vector and the TX antenna weight vector dimensions. This procedure creates a single dimension with the size of 144, because there were a total of 12 antenna weight vectors, each applied at the TX and the RX. The operation in question does not yield a loss of data, but it makes it more difficult to draw conclusions about the measured data, since the effects of TX and RX pointing angles are now mixed together. With this reduction to a 3-way tensor, there is one additional step needed before the results can be plotted. We consider each CIR and sum together the received power from each of the arriving paths that are present within the CIR. This reduces the delay dimension to only a single value and produces a matrix (equivalently, a 2-way tensor), which can be easily plotted.

\begin{figure}[!ht]
	\centering
	\includegraphics[scale=0.5]{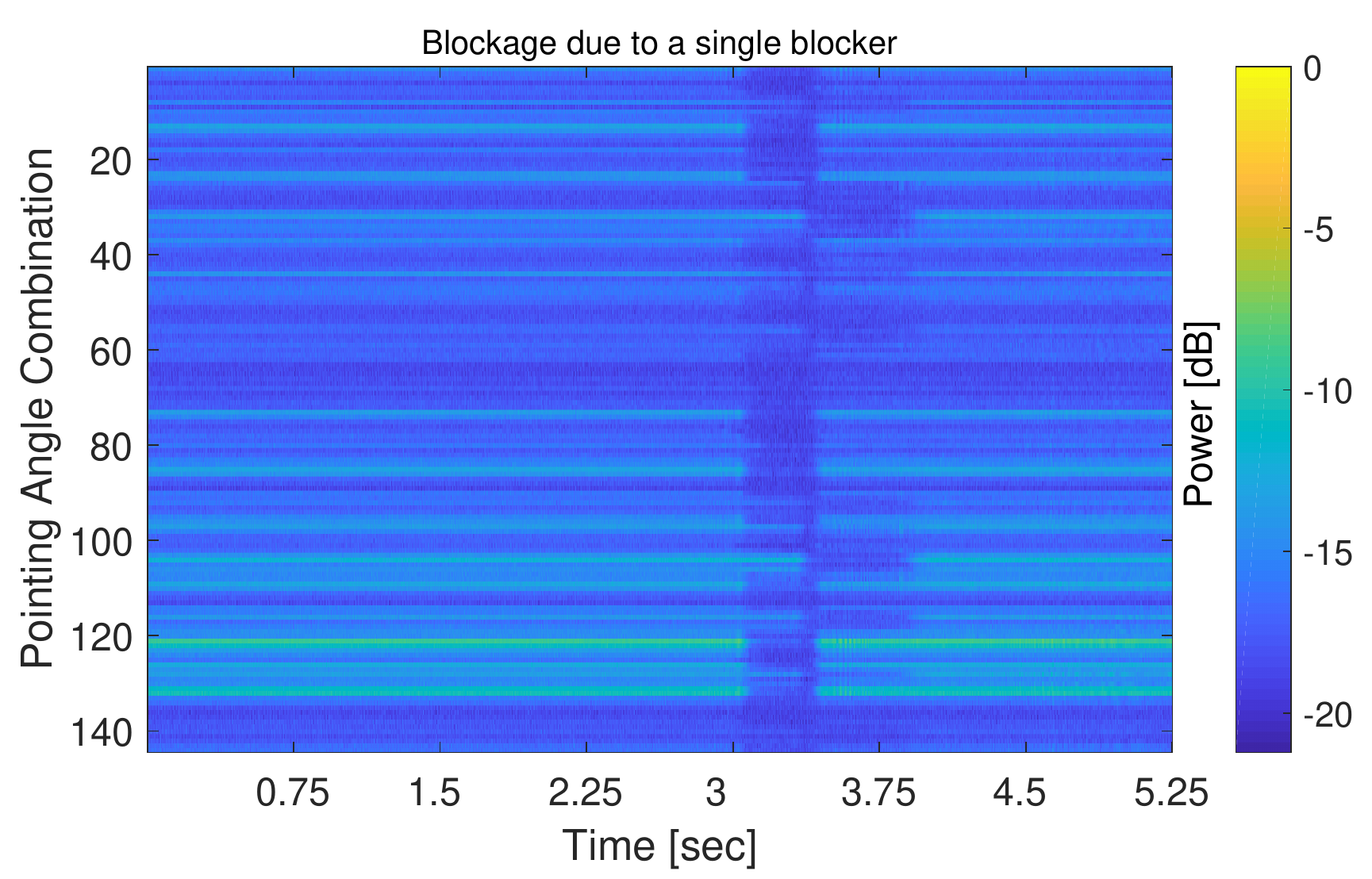}
	\includegraphics[scale=0.5]{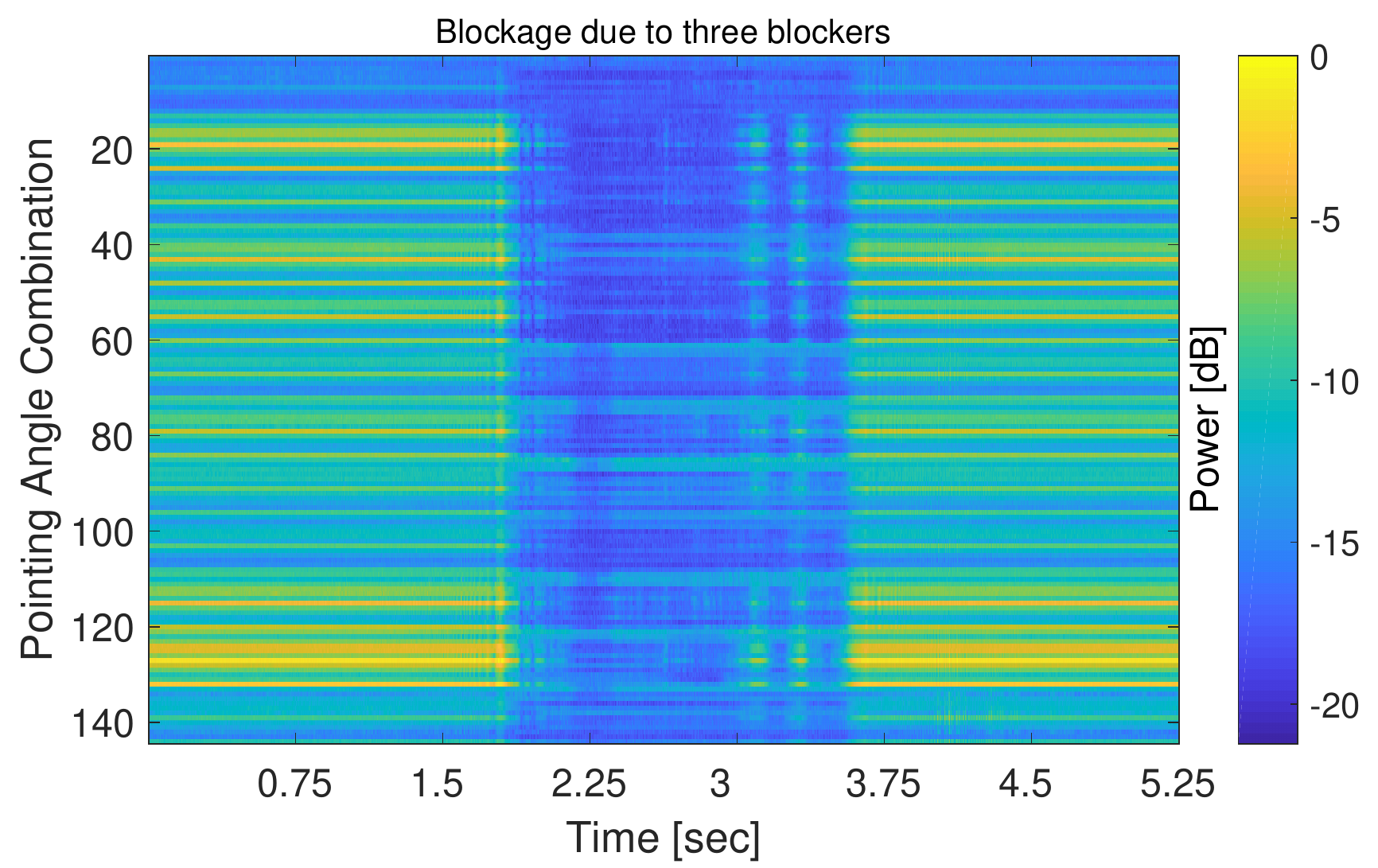}
	\caption{Received power for different numbers of human blockers. The single blocker measurement has a lower received power because the arrays were both pointing toward a wall and therefore fewer pointing angle combinations directed a substantial amount of energy to the LOS path.}
	\label{fig:density}
\end{figure}

Fig.~\ref{fig:density} demonstrates the results of two blockage measurements: (i) with a single person walking perpendicular to the LOS path and (ii) with three blockers following the paths as displayed in Fig.~\ref{fig:plan}. The single blocker measurement has the arrays facing towards the wall so that the NLOS component is relatively strong. There are two distinct blockage events visible in the single blocker measurement -- first the LOS component and then the NLOS component. There is an overlap between the two blockage events around the 3.5 second mark which has important implications for radio system design. If blockage mitigation is to be performed by switching between the LOS and NLOS path as each one experiences blockage, then an overlap similar to the one seen in Fig.~\ref{fig:density} could mean that there is simply no way to steer around this particular blockage event.

In the three-blocker measurement there is one very long blockage event on the LOS path due to the blocker who begins near the TX and walks toward the RX. This path is `unlucky' due to the fact that the blocker spends a large amount of time blocking the LOS path but is certainly something that could occur in a real use case. There is also a NLOS path that survives for most of the duration of the LOS blockage with the exception of some brief blockage events that are much shorter than the LOS blockage event. We can see two brief instances between 3 and 3.75 seconds where the LOS component appears to be coming out of blockage, but then becomes blocked again. This was likely caused by small variations in the blocker's path that briefly took him out of the way of the LOS between the TX and RX. These `false' recoveries might incur additional overhead if a radio transceiver switches back to the LOS component because it senses that it has been restored -- only to return to the blocked state shortly after. Overall, these measurements confirm that there is much to be gained from intelligent beam-tracking algorithms, which can select non-blocked paths.

\begin{figure}[!ht]
	\centering
	\includegraphics[scale=0.5]{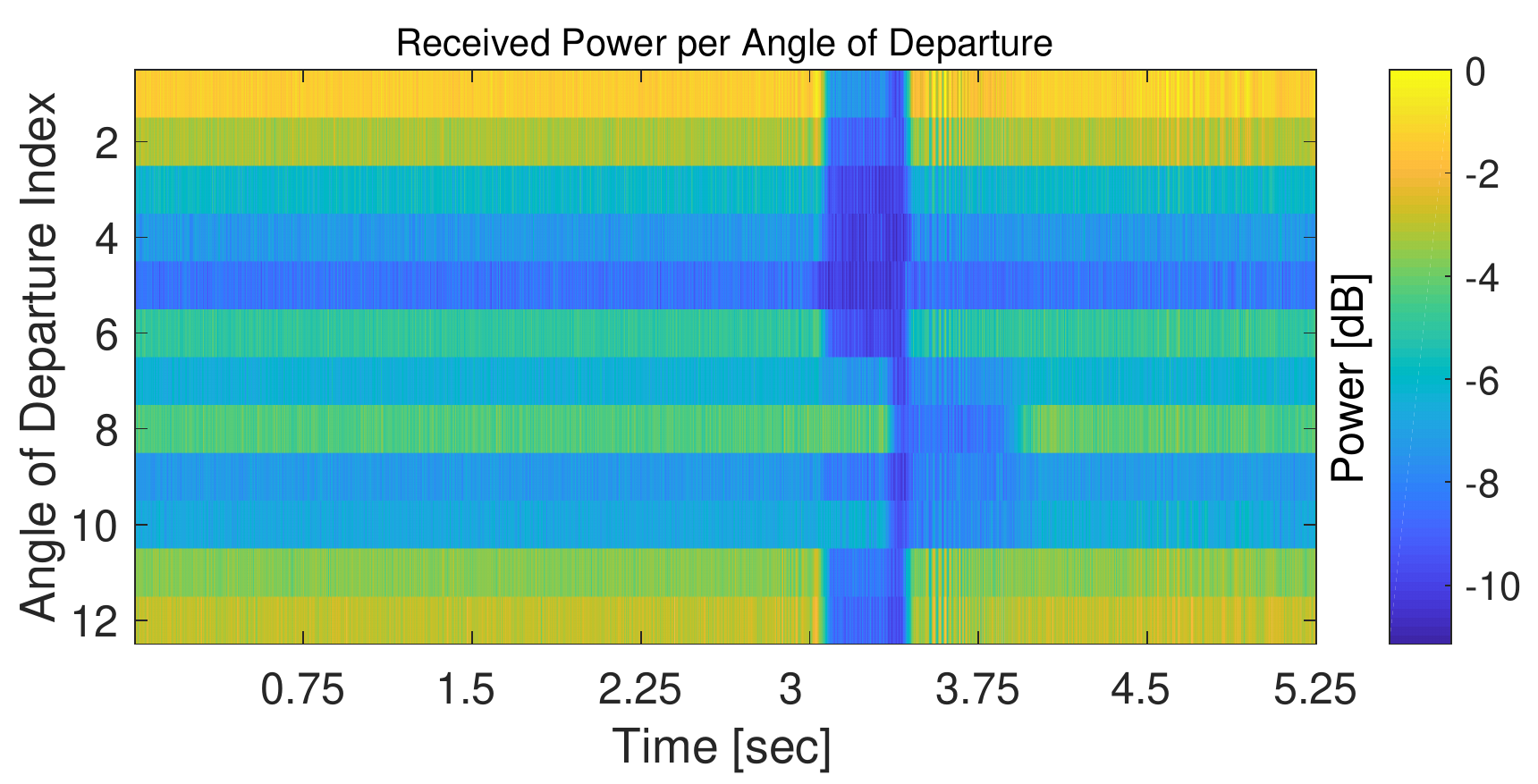}
	\caption{Received power for each AoD index (choosing the AoA with the largest received power) from a measurement with a single human blocker. Only one NLOS path is clearly visible.}
	\label{fig:aoaAndAod}
\end{figure}

Another way to visualize the results of our measurement is to plot the received power as a function of only the TX antenna weight vector (angle of departure) or the RX antenna weight vector (angle of arrival), while choosing the best pointing angle at the opposite end. An example of this plot for the single blocker measurement is introduced in Fig.~\ref{fig:aoaAndAod}. The latter is more intuitive than Fig.~\ref{fig:density}, which contains information from all of the possible combinations of the AoA and AoD.

By inspecting plots in the style of Fig. \ref{fig:aoaAndAod} or Fig. \ref{fig:density}, it is possible to label blockage events on the LOS path and any clearly visible NLOS paths. In this study, we seek to understand the temporal correlation between LOS and NLOS blockage. By examining all of the measured data collected during this campaign for the single blocker and multi-blocker measurements, we were able to conclude that there was never a moment where the LOS path and both NLOS paths reflected off the left and right walls would be simultaneously blocked. This outcome bodes well for 60 GHz applications, but as mentioned earlier more measurements in other environments and with different blocker trajectories are certainly needed in order to draw authoritative conclusions. For example, this observation may not hold if Blocker 3 in Fig. \ref{fig:plan} moved closer to the array. Further, this type of analysis is somewhat difficult to scale. As a way of automating the labeling process, we have developed a technique which uses a low-rank tensor decomposition to extract blockage profiles of the dominant paths in a single measurement.


\subsection{Low-Rank Tensor Decomposition}

A well-known method for reducing large datasets with multiple redundant or correlated entries is principal component analysis (PCA). PCA is suitable for a variety of problems and has been shown to be effective for interpreting the results of MIMO channel measurements at 6 GHz \cite{MIMOPCA}. PCA can easily be applied by performing the singular value decomposition (SVD) of the data matrix; the SVD is available in most standard linear algebra software packages which makes PCA a very accessible form of analysis. Because the SVD is only defined for matrices and not tensors, there are two ways to conduct the intended analysis. One way is to perform another unfolding of the data tensor as discussed earlier, but this time also interleaving the pointing angle combination along with the delay dimensions. The resulting matrix can then be used for standard PCA. However, as shown in~\cite{slezak60GHzBlockage}, this method is not as effective as other techniques, such as parallel factor analysis (PARAFAC), which generalize PCA to tensors. Intuitively, the PARAFAC model has fewer degrees of freedom than the equivalent PCA model which uses this interleaving process. If the PARAFAC model is sufficient to model the data, then these additional degrees of freedom can result in modeling noise or modeling system components in a redundant way \cite{ParafacTutorial}. The PARAFAC model expresses the data tensor as a sum of contributions from $L$ components, with each component defined as the outer product of its delay signature $\mathbf{d}$, its spatial signature $\mathbf{s}$, and its gain over time $\mathbf{g}$. In this way, PARAFAC provides a very elegant way of representing the results of a single measurement. The delay signature $d_{i\ell}$ tells us how strongly correlated multipath component $\ell$ is with delay $i$ in the power delay profile and therefore at what delay this multipath component is arriving. The spatial signature tells us how strongly correlated the $\ell^{th}$ component is with pointing angle combination $j$ and can therefore be used to determine where in the environment this path was located. Lastly, the gain trajectory $g_{k\ell}$ tells us how strong a particular component is at time $k$. Because all variation in the gain of each path in this experiment should be due to blockage, we can use this to obtain a blockage trace for each path in the channel.

\begin{figure}[!ht]
	\centering
	\includegraphics[scale=0.65]{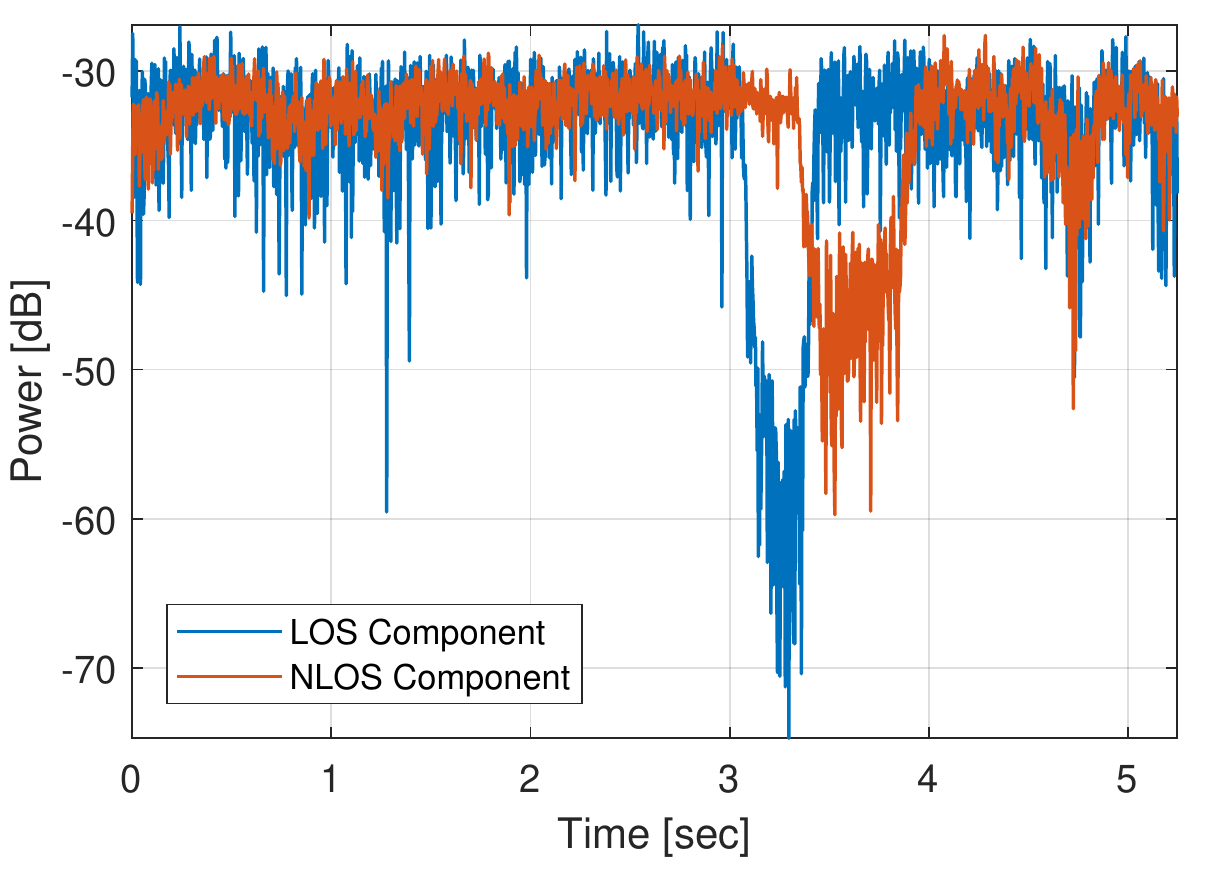}
	\caption{Blockage trajectories obtained from PARAFAC decomposition.}
	\label{fig:parafac}
\end{figure}

Fig.~\ref{fig:parafac} displays the two gain trajectories $g_{k1}$ and $g_{k2}$ obtained from a PARAFAC model (${L} = 2$) that was produced for the single blocker measurement shown in Fig.~\ref{fig:density}. These two curves have a straightforward physical interpretation as the time-varying received power of the two multipath components visible in the measurement. This very simple and usable form can be directly employed for the construction of statistical models. For instance, by fitting a piecewise linear model to the two blockage trajectories and then assigning states of `Blocked' or `Unblocked' to each path in every time slot, the Markov model demonstrated in~\cite{maccartneyRapidFadingCrowds} can be developed jointly for multiple paths instead of one path.

\section{Conclusions and Future Work}

	Characterizing fast channel dynamics due to blockage is vital for virtually every aspect of 
mmWave radio system design.  At the link layer, these models are needed for beam tracking, channel measurement
and reporting, and link adaptation. At the upper layers, they impact cell selection, handover 
and congestion control.  Models based on ray tracing, diffraction theory, or single directional measurements
are difficult to generalize to more complex scenarios.  In this work, we discussed the need for empirical studies of human-body blockage in mmWave communications. To address it, we developed and presented a novel methodology that can capture the spatial dynamics of mmWave channels using phased arrays. The system was demonstrated in a living room-type situation with multiple blockers.

The development of this tool can be regarded as a first step toward
the ultimate goal, which is constructing an accurate statistical framework that captures
the dynamic spatial and temporal nature of mmWave channels.  Beyond the work presented here, 
we anticipate that two further steps will be needed. 
First, raw data collected from our system is enormous and some form of dimensionality reduction is required.
We considered one possibility based on low-rank tensor decompositions, which can automatically extract
the dominant paths and their trajectories.  Second, we also need to fit a time-series type model to these trajectories.  In this case, we have multiple gains from different directional paths.  It is possible that 
the four-state models such as those used in IEEE 802.11 may apply, but these would have to be extended to handle
the joint dynamics across multiple paths.  This modeling challenge is an open and interesting one.

In addition to the statistical modeling of the existing data, our proposed methodology can also be applied to other
scenarios of interest.  For example, one can easily perform similar measurements in larger venues
or public spaces, such as cafes and typical hotspots where 5G systems are likely to be deployed. The PARAFAC method in this work requires that the only change in the channel over time is due to blockage, but it is possible to explore other formulations of this method or pre-processing techniques which make it suitable for other types of channel dynamics.

	\ignore{The research work presented in this article shows the necessity of the empirical studies of the human-body blockage in mmWave communications. The obtained results provide mechanisms for mitigation of the human blockage and shadowing effects and can be utilized in the future channel modeling and system design.

	One potential avenue of future work is to use our system and analysis techniques to explore hand blockage similar to \cite{qualcomm5GNRBlockage}.
	Can also perform measurements in larger venues. Experiments with blockers that are not knowingly part of the experiment, such as people in a crowded cafe. Experiments with multiple nodes in the network to study the effects of relaying.}

\bibliographystyle{IEEEtran}
\bibliography{commMagBib}

\begin{thebibliography}{10}
\providecommand{\url}[1]{#1}
\csname url@samestyle\endcsname
\providecommand{\newblock}{\relax}
\providecommand{\bibinfo}[2]{#2}
\providecommand{\BIBentrySTDinterwordspacing}{\spaceskip=0pt\relax}
\providecommand{\BIBentryALTinterwordstretchfactor}{4}
\providecommand{\BIBentryALTinterwordspacing}{\spaceskip=\fontdimen2\font plus
\BIBentryALTinterwordstretchfactor\fontdimen3\font minus
  \fontdimen4\font\relax}
\providecommand{\BIBforeignlanguage}[2]{{%
\expandafter\ifx\csname l@#1\endcsname\relax
\typeout{** WARNING: IEEEtran.bst: No hyphenation pattern has been}%
\typeout{** loaded for the language `#1'. Using the pattern for}%
\typeout{** the default language instead.}%
\else
\language=\csname l@#1\endcsname
\fi
#2}}
\providecommand{\BIBdecl}{\relax}
\BIBdecl

\bibitem{Shafi2017}
M.~Shafi \emph{et~al.}, ``{5G}: A tutorial overview of standards, trials,
  challenges, deployment, and practice,'' \emph{IEEE Journal on Selected Areas
  in Communications}, vol.~35, no.~6, pp. 1201--1221, June 2017.

\bibitem{fengLinkBlockage}
M.~Feng, S.~Mao, and T.~Jiang, ``Dealing with link blockage in {mmWave}
  networks: {D2D} relaying or multi-beam reflection?'' in \emph{2017 IEEE 28th
  Annu. Int. Symp. on Personal, Indoor, and Mobile Radio Communi. (PIMRC)}, Oct
  2017, pp. 1--5.

\bibitem{gustafson2011HumanShad60G}
C.~Gustafson and F.~Tufvesson, ``Characterization of 60 {GHz} shadowing by
  human bodies and simple phantoms,'' in \emph{2012 6th Eur. Conf. on Antennas
  and Propagation (EuCAP)}, April 2012, pp. 3084--3088.

\bibitem{jacob11ad60GModel}
M.~Jacob \emph{et~al.}, ``A ray tracing based stochastic human blockage model
  for the {IEEE} 802.11ad 60 {GHz} channel model,'' in \emph{2011 5th Eur.
  Conf. on Antennas and Propagation (EuCAP)}, April 2011, pp. 473--477.

\bibitem{rappaportSmallScalePropagation}
T.~S. Rappaport \emph{et~al.}, ``Small-scale, local area, and transitional
  millimeter wave propagation for {5G} communications,'' \emph{IEEE Trans. on
  Antennas and Propagation}, vol.~65, no.~12, pp. 6474--6490, Dec 2017.

\bibitem{peter11blockmeas}
M.~Peter \emph{et~al.}, ``Analyzing human body shadowing at 60 {GHz}:
  Systematic wideband {MIMO} measurements and modeling approaches,'' in
  \emph{2012 6th Eur. Conf. on Antennas and Propagation (EuCAP)}, April 2012,
  pp. 468--472.

\bibitem{3GPP38901}
\BIBentryALTinterwordspacing
3GPP, ``{Study on channel model for frequencies from 0.5 to 100 GHz (Release
  14)},'' Jan. 2018, version 14.3.0 (Accessed Aug. 24 2018). [Online].
  Available:
  \url{https://portal.3gpp.org/desktopmodules/Specifications/SpecificationDetails.aspx?specificationId=3173}
\BIBentrySTDinterwordspacing

\bibitem{METISRevised}
\BIBentryALTinterwordspacing
METIS, ``{METIS Channel Models},'' Jul. 2015, {Deliverable D1.4 v3 (Accessed
  Aug. 24 2018)}. [Online]. Available:
  \url{https://www.metis2020.com/wp-content/uploads/METIS_D1.4_v3.pdf}
\BIBentrySTDinterwordspacing

\bibitem{wangStochasticGeometry}
Y.~Wang \emph{et~al.}, ``Blockage and coverage analysis with mmwave cross
  street {BSs} near urban intersections,'' in \emph{2017 IEEE Int. Conf. on
  Commun. (ICC)}, May 2017, pp. 1--6.

\bibitem{collogneHumanActivity}
S.~Collonge, G.~Zaharia, and G.~E. Zein, ``Influence of the human activity on
  wide-band characteristics of the 60 {GHz} indoor radio channel,'' \emph{IEEE
  Trans. on Wireless Commun.}, vol.~3, no.~6, pp. 2396--2406, Nov 2004.

\bibitem{slezak60GHzBlockage}
C.~Slezak, A.~Dhananjay, and S.~Rangan, ``60 {GHz} blockage study using phased
  arrays,'' in \emph{2017 51st Asilomar Conf. on Signals, Syst., and Comput.},
  Oct 2017, pp. 1655--1659.

\bibitem{x60Paper}
S.~K. Saha \emph{et~al.}, ``{X60}: A programmable testbed for wideband 60 {GHz}
  {WLANs} with phased arrays,'' in \emph{Proc. Workshop on Wireless Network
  Testbeds, Experimental Evaluation \& CHaracterization}, Oct 2017, pp. 75--82.

\bibitem{maccartneyRapidFadingCrowds}
G.~R. MacCartney, T.~S. Rappaport, and S.~Rangan, ``Rapid fading due to human
  blockage in pedestrian crowds at {5G} millimeter-wave frequencies,'' in
  \emph{2017 IEEE Global Commun. Conf.}, Dec 2017, pp. 1--7.

\bibitem{MIMOPCA}
X.~Ma \emph{et~al.}, ``A {PCA}-based modeling method for wireless {MIMO}
  channel,'' in \emph{2017 IEEE Conf. on Comp. Commun. Workshops (INFOCOM
  WKSHPS)}, May 2017, pp. 874--879.

\bibitem{ParafacTutorial}
R.~Bro, ``{PARAFAC}. tutorial and applications,'' \emph{Chemometrics and
  Intelligent Laboratory Systems}, vol.~38, no.~2, pp. 149 -- 171, 1997.

\end{thebibliography}

\vspace{-0.8cm}

\begin{IEEEbiographynophoto}
    {Christopher Slezak} [GM]
 (chris.slezak@nyu.edu) received the B.S degree in Electrical and Computer Enginerring from Rutgers, the State University of New Jersey in 2014. He is currently a researcher at NYU WIRELESS working toward the Ph.D. degree at the NYU Tandon School of Engineering in Brooklyn NY. His research interests include wireless communications, millimeter wave channel measurements, and modeling of millimeter wave channel dynamics.
\end{IEEEbiographynophoto}

\vskip 0pt plus -1fil

\begin{IEEEbiographynophoto}
    {Vasilii Semkin}
(vasilii.semkin@tut.fi) is a Postdoctoral Researcher in the Laboratory of Electronics and Communications Engineering at Tampere University of Technology, Finland. He received B.Sc. degree from Saint-Petersburg State University of Aerospace Instrumentation, Russia, in 2009 and M.Sc. degree in 2011. He received Lic. Sc. (Tech.) in 2014 and D.Sc. (Tech.) in 2016 from Aalto University, School of Electrical Engineering. His interests comprise millimeter-wave communications, especially in the field of reconfigurable antennas and radio-wave propagation modeling.
\end{IEEEbiographynophoto}

\vskip 0pt plus -1fil

\begin{IEEEbiographynophoto}
    {Sergey Andreev} [SM]
(sergey.andreev@tut.fi) is a Senior Research Scientist in the Laboratory of Electronics and Communications Engineering at Tampere University of Technology, Finland. He received the Specialist degree (2006) and the Cand.Sc. degree (2009) both from St. Petersburg State University of Aerospace Instrumentation, St. Petersburg, Russia, and the Ph.D. degree (2012) from Tampere University of Technology. Sergey (co-)authored more than 150 published research works on wireless communications, energy efficiency, HetNets, cooperative communications, and M2M applications. 
\end{IEEEbiographynophoto}

\vskip 0pt plus -1fil

\begin{IEEEbiographynophoto}
    {Yevgeni Koucheryavy} [SM]
(yk@cs.tut.fi) is a Full Professor in the Laboratory of Electronics and Communications Engineering of Tampere University of Technology (TUT), Finland. He received his Ph.D. degree (2004) from TUT. He is the author of numerous publications in the field of communications. His current research interests include heterogeneous wireless communication networks and systems, the IoT, and nanocommunications. He is Associate Technical Editor of IEEE Communications Magazine and Editor of IEEE Communications Surveys and Tutorials.
\end{IEEEbiographynophoto}

\vskip 0pt plus -1fil

\begin{IEEEbiographynophoto}
	    {Sundeep Rangan} [F]
(sundeep.rangan@nyu.edu) is a Professor in Electrical and Computer Engineering
at New York University and Director of NYU Wireless.  He received
his Ph.D. from the University of California, Berkeley.  
He co-founded (with four others) Flarion Technologies, that developed
Flash-OFDM, the first cellular OFDM data system and pre-cursor to 4G LTE.  
Flarion was acquired by Qualcomm where Dr. Rangan was a  Director
of Engineering at Qualcomm before joining NYU.
\end{IEEEbiographynophoto}

\end{document}